\documentclass[a4paper,UKenglish,cleveref,autoref,thm-restate]{lipics-v2021}

\usepackage{listings}
\usepackage{booktabs}
\usepackage{adjustbox}
\usepackage[section]{placeins}
\graphicspath{{./}}

\providecommand{\Description}[1]{}

\title{Deterministic vs. Probabilistic Summarisation: An Empirical Trade-off Study in Design Pattern Centric Java Code}
\titlerunning{Deterministic vs. Probabilistic Summarisation}

\author{Najam Nazar}{Department of Software Systems and Cybersecurity, Monash University, Australia}{}{}{}
\author{Christoph Treude}{School of Computing and Information Systems, Singapore Management University, Singapore}{}{}{}

\authorrunning{N. Nazar and C. Treude}
\Copyright{Najam Nazar and Christoph Treude}



\category{}
\relatedversion{}
\acknowledgements{}

\EventLongTitle{The Empirical Software Engineering International Week 2026 (including the 20th International Symposium on Empirical Software Engineering and Measurement)}
\EventShortTitle{ESEIW 2026}
\EventAcronym{ESEIW}
\EventYear{2026}
\EventDate{October 4--9, 2026}
\EventLocation{Munich, Germany}
\begin{document}

\nolinenumbers

\maketitle

\begin{abstract}

\textbf{Background:}\enspace Automated code summarisation supports program comprehension and documentation, yet the relative strengths and limitations of \emph{deterministic} (heuristic-based) and \emph{probabilistic} (LLM-based) summarisation pipelines remain insufficiently understood.\\
\textbf{Aims:} This paper presents a controlled empirical comparison of these two paradigms for intent-oriented design pattern code summarisation.\\
\textbf{Method:} Using design-pattern-centric Java code as a structured evaluation testbed on 150 files from three open-source GitHub repositories covering nine common software design patterns, we compare a rule-based natural language generation (NLG) pipeline, a Software Word Usage Model (SWUM)-based approach, and a probabilistic pipeline based on the Mixtral large language model. We evaluate generated summaries against human references using automated similarity metrics, BERTScore and Cosine Similarity and complement these with rubric-based judgements produced by Llama~3 across five dimensions: accuracy, conciseness, adequacy, code-context awareness, and design-pattern fidelity. To strengthen inferential validity, we apply Wilcoxon signed-rank tests with effect sizes for pairwise metric comparisons, Friedman tests with post-hoc pairwise corrections for rubric-based rankings and Spearman correlation as a sensitivity analysis examining the consistency of the LLM-assisted rubric rankings.\\
\textbf{Results:} Across evaluations, probabilistic summaries exhibit stronger semantic alignment and richer contextual coverage, while deterministic approaches consistently produce more concise and fully reproducible outputs. Additional prompt-sensitivity and multi-run analyses show that LLM-based summaries remain subject to variability in length and phrasing, even under explicit constraints, though their relative performance trends are stable. %Our study spans 150 files from three open-source GitHub repositories covering nine common software design patterns.
\\
\textbf{Conclusions:} Overall, the results reveal a clear and persistent trade-off: probabilistic approaches are better suited to scenarios prioritising semantic depth and contextual accuracy, whereas deterministic pipelines remain preferable when brevity and strict reproducibility are required. These findings provide actionable guidance for researchers and tool builders selecting code summarisation techniques.
\end{abstract}

\section{Introduction}\label{sec:intro}
Code summarisation has received substantial attention in software engineering research and practice, with automated support traditionally provided through \emph{deterministic} pipelines based on heuristics, static analysis, and natural language generation (NLG), and more recently through \emph{probabilistic} approaches based on large language models (LLMs). Deterministic approaches offer predictable behaviour, explicit control over output structure, and reproducibility, whereas probabilistic approaches promise greater semantic coverage and flexibility at the cost of stochasticity and reduced controllability. While LLM-based techniques have seen rapid adoption, it remains unclear to what extent they outperform, complement, or trade off against well-established heuristic pipelines across dimensions such as semantic quality, conciseness, and reproducibility.

This paper addresses this gap through a controlled empirical comparison of deterministic and probabilistic approaches to intent-oriented code summarisation. We use design-pattern-centric code artefacts as a principled evaluation setting. Because design patterns impose well-defined roles, relationships, and interaction constraints, they provide an evaluation setting in which summaries must capture intent and architectural structure rather than surface-level functionality alone. This makes them particularly suitable for contrasting heuristic pipelines, whose outputs are tightly coupled to explicit structural features, with probabilistic models that infer intent through learned representations. Importantly, design patterns serve here as a controlled testbed for comparing summarisation paradigms, rather than as the primary object of study.

Our approach builds on a state-of-the-art Design Pattern Summarisation (DPS)~\cite{Nazar:2026} system and instantiates three automated summarisation pipelines that differ only in their summary generation strategy: (i) a deterministic, rule-based NLG approach, (ii) a deterministic Software Word Usage Model (SWUM)-based method and (iii) a probabilistic LLM-based pipeline using the Mixtral model. By holding preprocessing and pattern identification constant, this design isolates the effect of heuristic versus probabilistic generation. We benchmark the resulting summaries against human-written reference summaries drawn from the DPS corpus, which provides one manually written summary per file authored independently of this study, using automated similarity metrics (BERTScore and Cosine Similarity)~\cite{Nazar:2026}. In addition, we conduct rubric-based judgements with the \textit{Llama 3} model to assess five dimensions salient to developer utility, \textit{accuracy, conciseness, adequacy, code context awareness, and fidelity in capturing design pattern information}, which closely mirror the evaluation criteria used in prior work on design-pattern-centric code summarisation (e.g., Nazar et al.~\cite{Nazar:2026}).

Our results reveal a consistent trade-off between the two paradigms. Probabilistic, LLM-generated summaries achieve strong performance on accuracy, adequacy, code context awareness, and design-pattern fidelity, but are consistently less concise than deterministic pipelines. To further examine robustness and controllability, we conduct a series of prompt interventions. Specifically, we (a) remove the term \textit{concise} from the prompt to test whether conciseness emerges implicitly from the task description rather than from an explicit instruction, and (b) enforce explicit target lengths (20, 40, 60, and 80 words) under otherwise identical prompting. These experiments show that prompt refinements and explicit length constraints influence probabilistic outputs, but do not make them as predictable as deterministic summaries, highlighting a limited degree of controllability relative to heuristic approaches. Our evaluation is grounded in a corpus of 150 Java files drawn from three open-source GitHub repositories, covering nine common software design patterns.

%Our evaluation corpus comprises a subset of the existing DPS Corpus\footnote{\url{https://github.com/SamSike/DPS-Corpus}, accessed and verified on 15/01/26} drawn from two educational and one commercial open-source GitHub repositories: AbdurRKhalid\footnote{\url{https://github.com/SamSike/DPS-Corpus/tree/main/AbdurRKhalid}, accessed and verified on 15/01/26}, JamesZBL\footnote{\url{https://github.com/SamSike/DPS-Corpus/tree/main/JamesZBL}, accessed and verified on 15/01/26}, and randomly selected files from Spring-Framework\footnote{\url{https://github.com/spring-projects/spring-framework}, accessed and verified on 15/01/26} that specifically exhibit design patterns. The dataset contains 150 files, for which each method generates 150 summaries, enabling one-to-one comparisons with human-written references and systematic cross-method aggregation.

In short, this study contributes:

\begin{itemize}
    \item A controlled empirical comparison of heuristic, deterministic (NLG, SWUM) and probabilistic (LLM-based) approaches to intent-oriented code summarisation using a structured evaluation setting.
    \item A dual-vantage assessment combining automated similarity metrics (BERTScore, Cosine Similarity) with rubric-based judgements (Llama~3) over accuracy, conciseness, adequacy, code context awareness, and design-pattern fidelity.
    \item An analysis of prompt sensitivity and reproducibility that highlights a persistent conciseness trade-off and the non-deterministic nature of probabilistic summarisation relative to heuristic pipelines.
    \item An empirically grounded dataset spanning educational and commercial open-source repositories to support reproducible comparisons.
\end{itemize}

Together, these findings clarify when and how probabilistic summarisation provides measurable advantages over heuristic pipelines, and where traditional deterministic approaches remain preferable, particularly under strict conciseness or reproducibility requirements.

The structure of this paper is organised as follows: Section~\ref{sec:intro} introduces the research context and motivation. Section~\ref{sec:related_work} reviews related work in the domain of code summarisation, with particular emphasis on design patterns as an evaluation context and the role of large language models (LLMs). Section~\ref{sec:rq} formulates the research questions that guide our study. Section~\ref{sec:approach} details the proposed approach and methodology. Section~\ref{sec:results} presents the experimental results and evaluates them using statistical measures, LLM-based metrics, and human-generated summaries. Section~\ref{sec:implications} discusses implications of the study and Section~\ref{sec:threats} discusses internal and external threats to validity. Finally, Section~\ref{sec:conclusion} concludes the paper and outlines potential directions for future work.

\section{Related Work}\label{sec:related_work}
Early research in code summarisation predominantly relied on template-based and rule-driven natural language generation (NLG), coupling static analysis with heuristics over identifiers, parameters, and control structures to populate hand-crafted linguistic templates~\cite{Sridhara:2010,Sridhara:2011}. Foundational systems produced descriptive comments for Java methods via templates and linguistic modelling, and subsequent work extended these ideas to class-level summaries and statement selection guided by programmer salience~\cite{Sridhara:2010,Moreno:2013,Rodeghero:2014,Rodeghero:2015}. Similar efforts integrated lexical semantics with program structure through the Software Word Usage Model (SWUM), linking part-of-speech patterns in identifiers to program roles to improve phrasing and content selection~\cite{Malik:2009,Hill:2009,Pollock:2013}. Complementary context-aware approaches showed that analysing call sites and invocation context yields more informative summaries than code-only extraction, with user studies reporting statistically significant improvements over earlier automatic tools~\cite{McBurney:2014,McBurney:2016a}. Empirical eye-tracking studies further identified which tokens and statements developers attend to during summarisation, informing feature selection and improving the adequacy and readability of generated summaries~\cite{Rodeghero:2014,Rodeghero:2015,Karas:2024,Bansal:2025,McBurney:2016b}. Surveys synthesise these lines of work, noting strengths in interpretability but limitations in scalability and coverage across languages and domains, thereby motivating data-driven approaches~\cite{Zhang:2022,Zhu:2019,Zhang:2024}. Nazar et al.~\cite{Nazar:2023} introduce \textit{CodeLabeller}, a web-based system that crowdsources annotations of Java source files with design-pattern instances and natural-language summaries, producing a corpus of more than 1,000 files and reporting positive UEQ usability results from 25 software-engineering experts, thereby streamlining labelled-data collection for supervised learning in program comprehension.

\subsection{Large Language Models (LLMs) for Code Summarisation}
Neural encoder-decoder models with attention established the first end-to-end, data-driven baselines for code-to-text generation (e.g., Iyer et al.~\cite{Iyer:2016}) and were subsequently superseded by large pre-trained transformers specialised for code. Representative systems, PLBART~\cite{Ahmad:2021}, CodeT5/CodeT5+~\cite{Wang:2021,Wang:2023} and UniXcoder~\cite{Guo:2022}, consistently outperform earlier approaches on code summarisation across multiple languages and benchmarks~\cite{Ahmad:2021,Wang:2021,Wang:2023,Guo:2022,Husain:2019}. Evaluation practice commonly leverages large curated corpora such as \textit{CodeSearchNet}, which has become a de facto resource for code-to-text experiments and reproducible comparison~\cite{Husain:2019}.

Recent comprehensive studies in the LLM era report that advanced prompting strategies \textit{(few-shot, chain-of-thought, critique/expert)} do not uniformly outperform simple zero-shot prompting~\cite{Sun:2025b}, and that decoding hyper-parameters (e.g., temperature, top-$p$) materially affect performance across languages~\cite{Sun:2025b}. Notably, compact open models (e.g., CodeLlama-Instruct-7B) can outperform GPT-4 on specific summarisation facets such as design rationale or property assertions~\cite{Sun:2025b}. At the same time, traditional reference-based metrics (e.g., BLEU, ROUGE) correlate poorly with human judgements~\cite{Graham:2015,Mastropaolo:2024}, motivating the use of LLM-based evaluators; G-Eval with \textit{GPT-4} exhibits substantially stronger correspondence with human evaluation in summarisation than earlier metrics~\cite{Liu:2023,MicrosoftLearn:2024}. Privacy-preserving deployment has also been explored through knowledge distillation from \textit{GPT-3.5} to compact open models capable of near-parity summarisation on commodity GPUs~\cite{Su:2024}.

\subsection{Design Pattern-Oriented Summarisation}
While software design patterns are central to maintainability and communication, the bulk of automated research has focused on detecting or classifying pattern instances rather than summarising them in natural language~\cite{Yarahmadi:2020,Moreira:2022,ArcelliFontana:2012}. Detection approaches span ontology-based, rule-based, and machine-learning techniques (e.g., MARPLE-DPD, DPD\_F)~\cite{ArcelliFontana:2015,Fontana:2011,Alnusair:2014,Nazar:2026,Nazar:2022}. Multiple systematic reviews document this space and its limitations, including tool availability, accuracy variance, and the handling of pattern variants~\cite{Yarahmadi:2020,Moreira:2022,ArcelliFontana:2012}.

In contrast, DPS (Design Pattern Summarisation) introduced the first automated approach to summarising design-pattern instances using code features parsed with \textit{JavaParser}\footnote{\url{https://javaparser.org/} accessed and verified on 15/01/26} and NLG over a \textit{JSON} representation. Empirical results showed strong alignment with human-written summaries across \textit{ROUGE-L}, \textit{BLEU-4}, \textit{NIST}, and \textit{FrugalScore} metrics, with developers rating DPS higher for capturing summary context~\cite{Nazar:2026,Eddine:2022}. Despite this work, LLMs have not been systematically investigated for design-pattern summarisation. Recent LLM-based studies focus primarily on pattern recognition or identification rather than summarisation, underscoring an opportunity to examine both the strengths of probabilistic approaches, such as fluency and contextual reasoning, and their limitations, including prompt sensitivity and non-determinism, relative to deterministic pipelines~\cite{Pandey:2025,Pan:2025}.

\subsection{Our Contribution}
This paper investigates fundamental trade-offs between deterministic and probabilistic code summarisation in a deliberately structured, domain-specific evaluation setting: design-pattern-centric code artefacts. We compare rule-based NLG, SWUM and an LLM-based pipeline using Mixtral, evaluating their ability to capture architectural roles, interactions, and design intent. Through a combination of automated similarity metrics (e.g., BERTScore, Cosine Similarity) and rubric-based, model-assisted judgements of accuracy, conciseness, adequacy, code context awareness, and design-pattern fidelity, we identify where probabilistic approaches excel in semantic coverage and where deterministic pipelines remain advantageous.

\section{Research Questions}\label{sec:rq}
To explore the trade-offs between deterministic and probabilistic approaches to intent-oriented code summarisation, we formulate four complementary research questions. Rather than treating summarisation quality as a single aggregate outcome, these questions examine distinct dimensions of the trade-off, overall semantic quality, conciseness, domain sensitivity, and robustness, using design-pattern-centric code artefacts as a structured, domain-specific evaluation setting.

RQ1 examines overall summarisation quality. We ask whether probabilistic approaches provide measurable advantages over deterministic pipelines for intent-oriented summarisation in settings where summaries must capture architectural intent rather than surface-level functionality.

\textbf{RQ1: How do deterministic and probabilistic approaches compare for intent-oriented, design-pattern-centric code summarisation?}

This question is addressed using automated similarity metrics by comparing machine-generated summaries against a single human-written reference summary per file.

RQ2 focuses on the trade-off between semantic quality and conciseness. While probabilistic approaches may generate semantically richer summaries, this may come at the cost of brevity, which is an important practical constraint for documentation and developer-facing artefacts.

\textbf{RQ2: What trade-offs emerge between conciseness and semantic quality when using probabilistic summarisation approaches for design-pattern-centric code artefacts?}

We examine this question through rubric-based evaluations that jointly assess semantic quality and conciseness.

RQ3 investigates the role of domain structure. Because software design patterns differ in structural complexity and interaction style, we examine whether the observed trade-offs are consistent across patterns or influenced by pattern-specific characteristics.

\textbf{RQ3: Do the observed trade-offs between deterministic and probabilistic summarisation approaches vary across different software design patterns?}

RQ4 addresses robustness and controllability. Beyond average performance, developers and tool builders also care about the stability and predictability of generated summaries. We therefore examine how sensitive probabilistic summaries are to repeated runs and prompt-level interventions, relative to deterministic pipelines.

\textbf{RQ4: How stable are probabilistic summaries compared to deterministic ones when summarising design-pattern-centric code artefacts?}

We assume that design pattern identification is correct and treat detected pattern instances as given; evaluating pattern detection accuracy itself is outside the scope of this study.

\section{Methodology \& Approach}\label{sec:approach}

\subsection{Corpus}\label{subsec:corpus}
The evaluation corpus comprises a \textbf{subset} of the existing DPS Corpus~\cite{Nazar:2026}, selected to ensure coverage of both educational and industrial open-source Java code. Specifically, the corpus draws from two educational repositories — \textit{AbdurRKhalid} and \textit{JamesZBL} — and one commercial repository, the \textit{Spring Framework}, from which files were randomly sampled subject to the constraint that each file exhibits at least one identifiable design pattern. Together, the three repositories contribute 150 Java files spanning nine common software design patterns: Adapter, Visitor, Observer, Memento, Facade, Decorator, Abstract Factory, Factory Method, and Singleton. The educational repositories provide clearly structured, pattern-centric implementations suitable for controlled evaluation, while the Spring Framework files introduce industrial-scale complexity and heterogeneity, supporting a broader assessment of summarisation behaviour. For each file, all three pipelines generate one summary, yielding 150 summaries per method and enabling direct one-to-one comparisons against the human-written reference summaries provided in the DPS Corpus. This design supports both per-file pairwise statistical testing and systematic cross-method aggregation at the corpus level.

\begin{table}[!htbp]
  \caption{Example summary comparison across different methods}
  \label{tab:example_summaries}
  \centering
  \small
  \begin{adjustbox}{width=\textwidth}
  \begin{tabular}{p{2.6cm}p{5.2cm}p{2.3cm}p{3.7cm}}
    \toprule
    \textbf{File name} & \textbf{DPS\_NLG Summary} & \textbf{DPS\_SWUM Summary} & \textbf{DPS\_LLM Summary} \\
    \midrule
    Windows\-OS\-Factory & WindowsOSFactory acts as a factory for GUIFactory, which overrides createButton method of GUIFactory, createCheckBox method of GUIFactory. It is a public class that implements GUIFactory. The 2 methods of WindowsOSFactory are createButton (Button) and createCheckBox (CheckBox). No methods call createButton and createButton calls no methods. No methods call createCheckBox and createCheckBox calls no methods. & The WindowsOSFactory class acts as a factory in the Abstract Factory pattern and provides methods to create, create, check for managing factory. & The WindowsOSFactory class, implementing the GUIFactory interface, is responsible for creating Windows-specific GUI components. It collaborates with the Button and CheckBox classes to instantiate Windows-style versions of these components, adhering to the Abstract Factory design pattern. \\
    \midrule
    RoundPeg & RoundPeg acts as a target for adapter SquarePegAdapter, which adapts to RoundPeg. It is a public class. The only method of RoundPeg is getRadius (double). No methods call getRadius and getRadius calls no methods. & The RoundPeg class acts as a target in the adapter pattern and provides methods to get. & The RoundPeg class represents a peg with a radius, serving as a concrete component in the Adapter pattern. It provides a constructor and method to set and retrieve the radius. Collaborating with adapters, it enables compatibility with square pegs, adhering to the pattern's structural requirements. \\
    \midrule
    EventManager & EventManager acts as a publisher for observer EventListener, which is inherited by, which subscribes String parameter of eventType, EventListener parameter of listener, which unsubscribes String parameter of eventType, EventListener parameter of listener, which notifies String parameter of eventType, File parameter of file. It is a public class. The 3 methods of EventManager are subscribe (void), unsubscribe (void) and notify (void). No methods call subscribe and subscribe calls no methods. No methods call unsubscribe and unsubscribe calls no methods. No methods call notify and notify calls no methods. & The EventManager class acts as a publisher in the observer pattern and provides methods to notify for managing event. & The EventManager class, implementing the Observer pattern, manages event listeners. It maintains a map of event types to listener lists, collaborating with EventListener and File classes. It provides methods to subscribe, unsubscribe, and notify listeners about events, facilitating event-driven communication. \\
    \bottomrule
  \end{tabular}
  \end{adjustbox}
\end{table}

\subsection{DPS}
The \textbf{Design Pattern Summariser (DPS)}~\cite{Nazar:2026} provides an automated pipeline for generating contextual summaries of software design patterns directly from source code. In this work, DPS serves as a common preprocessing and representation framework that enables a controlled comparison of different summarisation paradigms. DPS leverages \textit{JavaParser} to extract code features, converts structural information into a \textit{JSON} representation, and employs a Natural Language Generation (NLG) component to produce human-readable summaries. The shared infrastructure allows us to isolate the effect of the summary generation strategy, deterministic versus probabilistic, while holding all upstream analysis constant. Below, we briefly describe the DPS workflow and then detail the alternative generation pipelines used in our study. In short, DPS generates design-pattern summaries through a three-stage pipeline: (i) preprocessing, in which JavaParser constructs an Abstract Syntax Tree from Java source code and extracts up to 20 structural and behavioural code features (e.g., class names, inheritance relationships, method dependencies), consolidating these into a JSON representation; (ii) pattern identification, in which a Pattern Checker module identifies design patterns from the extracted features and enriches the JSON with pattern-specific roles and relationships (e.g., subscriber–publisher structure in the Observer pattern); and (iii) summarisation, in which the enriched JSON is transformed into natural language text. In this paper, Stages i and ii are held constant across all three pipelines — DPS\_NLG, DPS\_SWUM, and DPS\_LLM — and only the summarisation stage varies, isolating the effect of generation strategy as the sole independent variable. Full details of the DPS architecture are available in the original DPS publication~\cite{Nazar:2026}. In subsequent sections we discuss SWUM and LLM based DPS pipelines.

\subsubsection{SWUM driven DPS Summarisation}\label{subsec:swum}
The Software Word Usage Model (SWUM) is a linguistically motivated, fully deterministic approach designed to improve program comprehension by exploiting natural language information embedded in source code identifiers~\cite{Pollock:2013}. Unlike summarisation techniques that rely primarily on structural representations, SWUM models semantic relationships between words in identifiers to infer the intent and functionality of code elements. By analysing verbs, objects, and thematic roles in method and variable names, SWUM enables the generation of summaries that more closely reflect how developers conceptualise program behaviour. For example, from a method name such as \textit{buildQueryForTrace}, SWUM infers that the method constructs a query for a trace, capturing both the action and its target.

SWUM has demonstrated extensibility across programming languages, initially developed for Java and later adapted for C++ through structural and syntactic modifications~\cite{Malik:2009}. Its summarisation pipeline typically involves parsing identifiers, extracting linguistic features, and applying rules to generate natural language phrases that describe code functionality. Empirical studies show that incorporating linguistic cues improves summary quality relative to purely structural approaches. In our implementation, we develop a Java-based SWUM pipeline that operates on the same code features and JSON representation as \textbf{DPS\_NLG}, ensuring comparability across summarisation strategies.

%\subsubsection{DPS\_LLM}\label{subsec:llm}
\subsubsection{LLM driven DPS Summarisation}\label{subsec:llm}
To examine how probabilistic, LLM-based generation behaves within the same structured framework, we employ the \textbf{Mixtral:8x22B} model to generate design-pattern-centric summaries. Mixtral:8x22B is a Sparse Mixture of Experts (SMoE) language model developed by Mistral AI, designed for efficient large-scale text generation and summarisation tasks~\cite{Mistral:2024}. Its architecture consists of eight expert feedforward blocks per layer, with a router dynamically selecting two experts per token, activating approximately 39 billion parameters out of a total of 141 billion during inference~\cite{Mistral:2024,APXML:2024}. This design reduces computational overhead while maintaining strong performance across multilingual and reasoning benchmarks~\cite{Mistral:2024,AINews:2024}. The model supports an extended context window of up to 64,000 tokens, enabling it to process rich structural inputs without truncation. Prior evaluations report that Mixtral outperforms models such as \textit{LLaMA 2 70B} and \textit{GPT-3.5} on tasks requiring long-context understanding and structured text generation~\cite{Jiang:2024}.

We configure the model with a temperature of \textbf{zero} and a maximum token limit of \textbf{512}. Although temperature zero reduces sampling variability, hosted LLM inference may still produce small output differences across repeated calls because of implementation-level non-determinism, model-serving infrastructure, or API-side changes outside our control. Summaries are generated by issuing a system prompt to the Mixtral API via OpenRouter at \url{https://openrouter.ai/api/v1/chat/completions}. Following the prompting strategy proposed by Schindler et al.~\cite{Schindler:2025}, we use the JSON based prompt shown in prompt box below. The default configuration instructs the model to generate a single-paragraph summary of no more than 50 words.

\begin{lstlisting}[caption={System prompt configuration for the Senior Analyst agent},label={lst:system_prompt},basicstyle=\ttfamily\footnotesize,breaklines=true]
"system_prompt": {
    "alias": "SENIOR_ANALYST_50_WORDS",
    "content": "You are a senior software analyst who writes concise, factual summaries of Java classes. Use only the supplied structural facts to describe each class in a single paragraph no longer than 50 words. Highlight the class responsibilities, key collaborators, and any explicit design-pattern roles. Do not invent behaviour or reference missing information. Write in third person with an objective tone."
},
\end{lstlisting}

%\begin{lstlisting}[caption={User prompt template for the textual summaries},label={lst:user_prompt},basicstyle=\ttfamily\footnotesize,breaklines=true]
%user_prompt": {
%      "alias": "CLASS_SUMMARY_STRUCTURED",
%      "sections": [
%        "Project: {project_name}",
%        "Class: {class_name} ({class_kind}, modifiers: {modifiers})",
%        "Extends: {extends_types}",
%        "Implements: {implements_types}",
%        "Fields (total {field_count}): {field_signatures}",
%        "Constructors (total {constructor_count}): {constructor_signatures}",
%        "Methods (total {method_count}): {method_summaries}",
%        "Interactions: {interaction_notes}",
%        "Design pattern insights:\n{pattern_insights}",
%        "\nTask: Produce a single-paragraph summary (<=50 words) that stresses the class responsibility, its collaborators, and the provided design-pattern context. Avoid repeating raw bullet text."
%      ]
%    },
%\end{lstlisting}

As specified in the prompt, the goal is to generate concise summaries with a target length of at most 50 words. Table~\ref{tab:example_summaries} illustrates example outputs produced by the \textbf{DPS\_NLG}, \textbf{DPS\_SWUM}, and \textbf{DPS\_LLM} pipelines for three representative Java files.

\section{Results and Evaluation}\label{sec:results}
This section reports the results of our evaluation and addresses the four research questions defined in Section~\ref{sec:rq}.

\subsection{RQ1: How do deterministic and probabilistic approaches compare for intent-oriented, design-pattern-centric code summarisation?}\label{subsec:rq1}

To compare deterministic and probabilistic summarisation approaches, we evaluate system outputs against human-written reference summaries using two automated semantic similarity metrics: BERTScore and Cosine Similarity. We also applied the \textbf{Wilcoxon} pairwise test on these methods. All three DPS variants, \textbf{DPS\_NLG}, \textbf{DPS\_SWUM} and \textbf{DPS\_LLM}, are evaluated under identical conditions. 

Table~\ref{tab:comparison} reports average Cosine Similarity and BERTScore (Precision, Recall, and F1-Score) for each method.

\begin{table}[!htbp]
  \caption{Comparison of summaries generated through DPS variants with human summaries using Cosine Similarity and BERTScore.}
  \label{tab:comparison}
  \centering
  \small
  \begin{adjustbox}{width=\textwidth}
  \begin{tabular}{lcccc}
    \toprule
    \textbf{Metric} & \textbf{DPS\_NLG} & \textbf{DPS\_SWUM} & \textbf{DPS\_LLM} & \textbf{DPS\_LLM (non-concise)} \\
    & \textbf{vs Human} & \textbf{vs Human} & \textbf{vs Human} & \textbf{vs Human} \\
    \midrule
    Average Cosine Similarity & 0.1657 & 0.2532 & 0.3351 & 0.3275 \\
    Average BERTScore Precision & 0.8361 & 0.8932 & 0.8702 & 0.8685 \\
    Average BERTScore Recall & 0.8596 & 0.8589 & 0.8901 & 0.8901 \\
    Average BERTScore F1 & 0.8473 & 0.8754 & 0.8799 & 0.8790 \\
    \bottomrule
  \end{tabular}
  \end{adjustbox}
\end{table}

Across all metrics, \textbf{DPS\_LLM (concise)}, generated using the default prompt with an explicit $\leq$50-word constraint, achieves the strongest overall performance. It attains the highest average Cosine Similarity (0.3351) and the highest BERTScore F1 (0.8799), indicating closer semantic alignment and more comprehensive coverage relative to human-written summaries. The \textbf{DPS\_LLM (non-concise)} variant, generated by removing the conciseness instruction from the prompt while keeping all other settings identical, performs slightly worse across most metrics, suggesting that increased verbosity does not improve semantic fidelity and may dilute Precision.

The \textbf{DPS\_SWUM} pipeline ranks third overall. It achieves the highest BERTScore Precision (0.8932), indicating conservative but accurate content selection, yet exhibits lower Cosine Similarity (0.2532), suggesting more limited semantic breadth. \textbf{DPS\_NLG} performs consistently worst across all metrics, with the lowest Cosine Similarity (0.1657), reflecting the limitations of template-based generation for capturing higher-level intent.

Two-sided Wilcoxon signed-rank tests conducted on 150 matched pairs revealed that DPS\_LLM significantly outperformed DPS\_NLG on both Cosine Similarity (median difference = $-0.165$) and BERTScore F1 (median difference = $-0.026$), and similarly outperformed DPS\_SWUM on Cosine Similarity (median difference = $-0.085$). However, the difference between DPS\_SWUM and DPS\_LLM on BERTScore F1 was not statistically significant (median difference = $0.001$), indicating that DPS\_SWUM's conservative but precise content selection yields contextualised token-level alignment comparable to the probabilistic approach on this metric.

\begin{figure}[!htbp]
  \centering
  \includegraphics[width=0.95\textwidth]{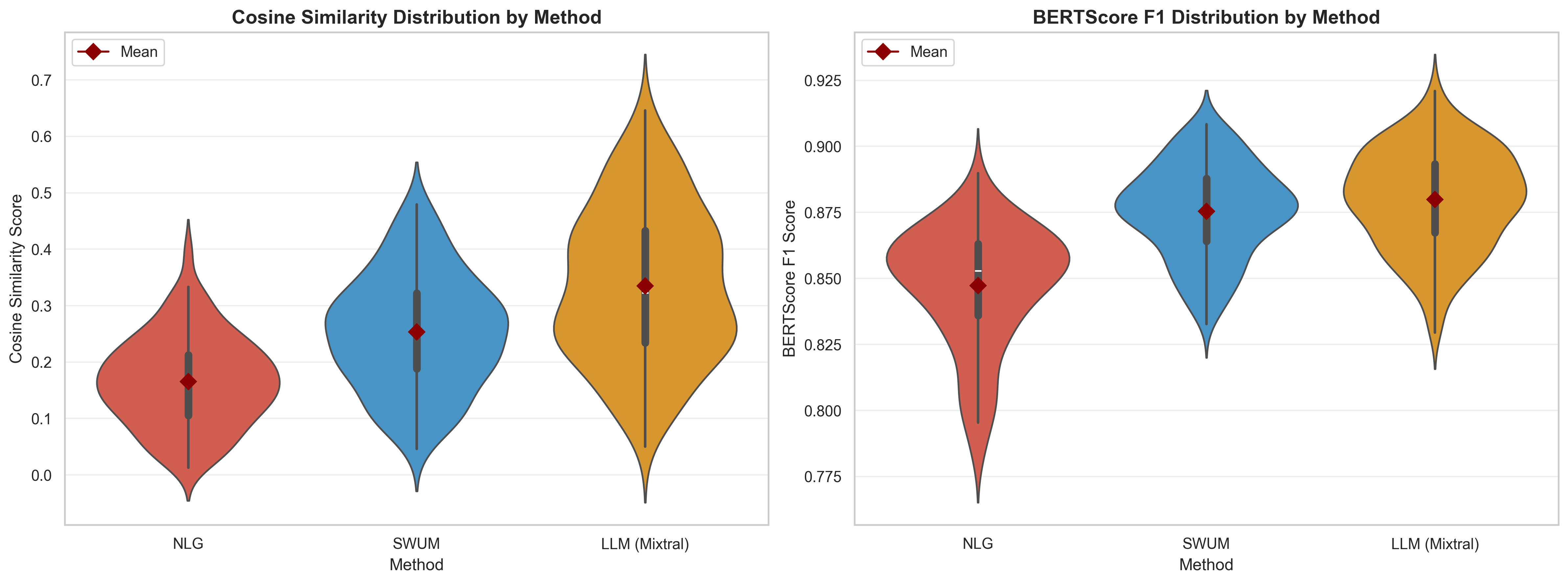}
  \caption{Violin plots comparing score distributions across three methods (NLG, SWUM, and LLM). Left: Cosine Similarity; right: BERTScore F1. Width indicates density, and red diamonds mark the mean.}
  \Description{Two violin plots comparing score distributions for NLG, SWUM, and LLM methods. The left panel shows Cosine Similarity and the right panel shows BERTScore F1.}
  \label{fig:violin_plots}
\end{figure}

To complement aggregate metrics, we visualise score distributions using violin plots (Figure~\ref{fig:violin_plots}), focusing on \textbf{DPS\_NLG}, \textbf{DPS\_SWUM}, and the concise \textbf{DPS\_LLM} variant.

\FloatBarrier

The violin plots reveal consistent ordering across both metrics. \textbf{DPS\_LLM} exhibits the highest mean scores and a broader upper tail, indicating stronger best-case performance and greater semantic coverage. \textbf{DPS\_SWUM} shows a narrower distribution with moderate variability, while \textbf{DPS\_NLG} produces consistently lower scores with limited spread. The broader distribution for LLM-based summaries reflects higher expressiveness, whereas deterministic methods exhibit tighter but lower-performing distributions.

%Overall, these results indicate that probabilistic summarisation provides superior semantic alignment with human-written summaries for intent-oriented code, while deterministic approaches trade semantic breadth for tighter control and consistency.

%\begin{quote}
\textbf{RQ1:} Probabilistic summarisation consistently achieves higher semantic alignment with human-written summaries than deterministic pipelines. Across automated similarity metrics, LLM-based summaries more closely approximate human descriptions of design intent and context, while deterministic approaches lag in semantic coverage.
%\end{quote}

\subsection{RQ2: What trade-offs emerge between conciseness and semantic quality when using probabilistic summarisation approaches for design-pattern-centric code artefacts?}\label{subsec:rq2}

%To characterise the trade-off between conciseness and semantic quality, we first analyse summary length distributions across the three DPS variants. Table~\ref{tab:wordstats} reports average, minimum, and maximum word and character counts.

%\begin{table}[!htbp]
%  \caption{Average, minimum, and maximum words and characters per summary}
%  \label{tab:wordstats}
%  \centering
%  \small
%  \begin{tabular}{lrrrrrr}
%    \toprule
%    \textbf{Method} & \textbf{Avg} & \textbf{Min} & \textbf{Max} & \textbf{Avg} & \textbf{Min} & \textbf{Max} \\
%     & \textbf{Words} & \textbf{Words} & \textbf{Words} & \textbf{Chars} & \textbf{Chars} & \textbf{Chars} \\
%    \midrule
%    DPS\_LLM  & 42.51 & 28 & 57  & 268.6  & 168 & 405 \\
%    DPS\_SWUM & 13.61 & 3  & 31  & 76.31  & 13  & 173 \\
%    DPS\_NLG  & 57.51 & 12 & 151 & 372.79 & 54  & 934 \\
%    \bottomrule
%  \end{tabular}
%\end{table}

%The results show clear differences in verbosity. \textbf{DPS\_SWUM} produces the most compact summaries, averaging 13.61 words, while \textbf{DPS\_NLG} generates the longest outputs, averaging 57.51 words with high variance, due to exhaustive template expansion over detected classes, methods, and relationships rather than selective content compression. \textbf{DPS\_LLM} occupies an intermediate position, with an average of 42.51 words and a relatively narrow range (28--57 words), reflecting steadier verbosity but limited brevity even under explicit length constraints.

To assess whether increased verbosity corresponds to improved semantic quality, we evaluate summaries across five qualitative criteria: \textit{Accuracy}, \textit{Conciseness}, \textit{Adequacy}, \textit{Code Context}, and \textit{Design Pattern fidelity}. We use an LLM-assisted ranking procedure based on the \textit{Llama 3 70B} model,\footnote{\url{https://www.llama.com/models/llama-3/} accessed and verified on 15/01/26} selected for its strong instruction-following performance and prior use as an automated evaluator in summarisation tasks. Importantly, the judge model (Llama 3) is distinct from the model used for summary generation (Mixtral), avoiding self-evaluation and reducing the risk of model-specific bias. A single judge model is used consistently across all evaluations to ensure comparability. %To mitigate bias and avoid model conditioning on method identity, summaries are randomly labelled as \textbf{A}, \textbf{B}, and \textbf{C} before ranking.
For each criterion, the model ranks the three summaries relative to the corresponding human-written reference. For each criterion, the judge produces a strict ranking of the three summaries; ties are not permitted.

% \begin{table}[!htbp]
%   \caption{Performance of all three summaries across five criteria relative to human references}
%   \label{tab:rankings}
%   \centering
%   \small
%   \begin{adjustbox}{width=\textwidth}
%   \begin{tabular}{lcccc}
%     \toprule
%     \textbf{Metric} & \textbf{Corpus A} & \textbf{Corpus B} & \textbf{Corpus C} & \textbf{Total} \\
%     & \textbf{(1st/2nd/3rd)} & \textbf{(1st/2nd/3rd)} & \textbf{(1st/2nd/3rd)} & \textbf{(1st/2nd/3rd)} \\
%     \midrule
%     ACCURACY & 3 / 82 / 64 & 145 / 0 / 4 & 1 / 67 / 81 & 149 / 149 / 149 \\
%     CONCISENESS & 81 / 19 / 49 & 67 / 10 / 72 & 1 / 47 / 102 & 149 / 149 / 149 \\
%     ADEQUACY & 5 / 102 / 43 & 144 / 1 / 5 & 1 / 47 / 102 & 150 / 150 / 150 \\
%     CODE CONTEXT & 3 / 93 / 53 & 143 / 1 / 5 & 3 / 55 / 91 & 149 / 149 / 149 \\
%     DESIGN PATTERN & 6 / 95 / 46 & 133 / 1 / 13 & 8 / 51 / 88 & 147 / 147 / 147 \\
%     \bottomrule
%   \end{tabular}
%   \end{adjustbox}
% \end{table}

\begin{table}[!htbp]
  \caption{Performance of all three summaries across five criteria relative to human references}
  \label{tab:rankings}
  \centering
  \small
  \begin{adjustbox}{width=\textwidth}
  \begin{tabular}{lcccc}
    \toprule
    \textbf{Metric} & \textbf{Corpus NLG} & \textbf{Corpus LLM} & \textbf{Corpus SWUM} & \textbf{Total} \\
    & \textbf{(1st/2nd/3rd)} & \textbf{(1st/2nd/3rd)} & \textbf{(1st/2nd/3rd)} & \textbf{(1st/2nd/3rd)} \\
    \midrule
    ACCURACY & 3 / 82 / 64 & 145 / 0 / 4 & 1 / 67 / 81 & 149 / 149 / 149 \\
    CONCISENESS & 81 / 19 / 49 & 67 / 10 / 72 & 1 / 47 / 102 & 149 / 149 / 149 \\
    ADEQUACY & 5 / 102 / 43 & 144 / 1 / 5 & 1 / 47 / 102 & 150 / 150 / 150 \\
    CODE CONTEXT & 3 / 93 / 53 & 143 / 1 / 5 & 3 / 55 / 91 & 149 / 149 / 149 \\
    DESIGN PATTERN & 6 / 95 / 46 & 133 / 1 / 13 & 8 / 51 / 88 & 147 / 147 / 147 \\
    \bottomrule
  \end{tabular}
  \end{adjustbox}
\end{table}

Across the LLM-assisted rubric judgements, \textbf{Corpus LLM} is ranked as the strongest performer. It receives the highest number of first-place rankings in \emph{Accuracy} (145/149), \emph{Adequacy} (144/150), \emph{Code Context} (143/149), and \emph{Design Pattern fidelity} (133/147), yielding the highest overall score on the 3--2--1 ranking scale (mean $\approx 2.72$). However, this advantage does not extend to \emph{Conciseness}, where \textbf{Corpus NLG} achieves the most first-place rankings (81/149), despite underperforming on other criteria.

\textbf{Corpus SWUM} most frequently ranks third across criteria, particularly for \emph{Accuracy}, \emph{Adequacy}, and \emph{Code Context}. While it produces the shortest summaries overall, this brevity often coincides with reduced semantic coverage. The \emph{Design Pattern} criterion contains slightly fewer observations (147) due to a small number of malformed LLM ranking outputs, which were discarded and do not materially affect the results.

Friedman tests conducted per criterion across 150 class files revealed statistically significant differences among the three methods on all five criteria with post-hoc pairwise tests confirming that DPS\_LLM significantly outperformed both deterministic methods on all semantic criteria. The DPS\_NLG significantly outperformed DPS\_SWUM on \textbf{Conciseness} (adjusted p < 0.001), and a cross-criteria Friedman test further confirmed that this ranking pattern was itself consistent across all five criteria (W = 0.840), with the sole exception that DPS\_NLG held the conciseness advantage over DPS\_LLM, though this difference did not survive Bonferroni correction (adjusted p = 0.131).

\subsubsection{LLM Judge Consistency Analysis}\label{subsubsec:bias}

To examine whether the Llama 3 judge applies the rubric criteria consistently, we compute
\textit{Spearman rank correlations} between criterion-level rankings and the aggregate ranking outcomes. This analysis does not eliminate the possibility of evaluator bias, but it provides a sensitivity check on whether the five criteria behave similarly or capture distinct aspects of summary quality. Spearman correlation analysis conducted across 450 observations per criterion showed that criterion-level rankings were significantly and positively associated with aggregate ranking outcomes for all five criteria ($\alpha$ = 0.05), with Accuracy ($\rho$ = 0.912, p < 0.001), Adequacy ($\rho$ = 0.915, p < 0.001), Code Context ($\rho$ = 0.901, p < 0.001), and Design Pattern ($\rho$ = 0.881, p < 0.001) each exhibiting very strong monotonic associations. The Conciseness factor demonstrates a comparatively moderate association ($\rho$ = 0.400, p < 0.001), suggesting that the four semantic criteria contribute strongly to the aggregate rankings, whereas Conciseness captures a more distinct aspect of summary quality.

%Taken together, these results indicate that improved semantic quality in probabilistic summarisation is consistently associated with reduced conciseness, while deterministic approaches achieve brevity at the expense of semantic breadth. Importantly, this trade-off persists despite explicit prompt constraints on output length, suggesting that conciseness is a systematic limitation rather than a prompt-level artefact.

%\begin{quote}
%\textbf{RQ2:} The gains in semantic quality achieved by probabilistic summarisation come at a clear cost in conciseness. While LLM-based summaries dominate in accuracy, adequacy, code context, and design-pattern fidelity, they are consistently outperformed by deterministic approaches on brevity, even under explicit length constraints.
%\end{quote}

\subsection{RQ3: Do the observed trade-offs between deterministic and probabilistic summarisation approaches vary across different software design patterns?}\label{subsec:rq3}

To examine whether the observed trade-offs depend on pattern-specific characteristics, we analyse summary length statistics across different software design patterns. Table~\ref{tab:patterns} reports average word counts and related descriptive statistics for summaries generated for each pattern.

\begin{table}[!htbp]
  \caption{Average word count statistics across different design patterns}
  \label{tab:patterns}
  \centering
  \small
  \begin{adjustbox}{width=\textwidth}
  \begin{tabular}{lccccccc}
    \toprule
    \textbf{Pattern} & \textbf{Count} & \textbf{AvgWords} & \textbf{Median} & \textbf{StdWords} & \textbf{AvgChars} & \textbf{AvgWord/} & \textbf{AvgTTR} \\
    & & & \textbf{Words} & & & \textbf{Sent} & \\
    \midrule
    Adapter & 14 & 49.14 & 40 & 4.05 & 254.79 & 14.67 & 0.80 \\
    Visitor & 19 & 40.58 & 41 & 4.18 & 258.05 & 17.07 & 0.83 \\
    Observer & 14 & 41.43 & 40 & 4.57 & 261.50 & 15.30 & 0.83 \\
    Memento & 10 & 44.30 & 45.5 & 4.92 & 275.30 & 14.02 & 0.79 \\
    Facade & 18 & 42.94 & 43.5 & 6.37 & 285.44 & 12.80 & 0.85 \\
    Decorator & 14 & 42.21 & 42.21 & 3.70 & 271.57 & 16.46 & 0.79 \\
    Abstract Factory & 29 & 41.69 & 41.69 & 4.46 & 266.66 & 16.28 & 0.82 \\
    Factory Method & 20 & 42.70 & 42.70 & 5.63 & 266.05 & 15.56 & 0.80 \\
    Singleton & 12 & 44.30 & 43.5 & 3.98 & 284.30 & 15.85 & 0.83 \\
    \bottomrule
  \end{tabular}
  \end{adjustbox}
\end{table}

Across the nine design patterns, summary lengths are highly consistent, clustering in the low 40-word range rather than approaching the upper bound of 50 words. Median values closely track the means, and variability remains moderate (\textit{StdWords} $\approx$ 3.7--6.37), indicating stable summarisation behaviour across patterns.

By average word count, \emph{Adapter} produces the longest summaries (49.14 words), while \emph{Visitor} (40.58), \emph{Observer} (41.43), and \emph{Abstract Factory} (41.69) are the most concise. Character counts show minor variation, with \emph{Facade} exhibiting the largest footprint (approximately 285 characters) and \emph{Adapter} the smallest (approximately 255 characters), reflecting differences in average word length rather than structural divergence.

Sentence-level characteristics are similarly stable. Average words per sentence range narrowly from 12.8 (\emph{Facade}) to 17.07 (\emph{Visitor}), corresponding to roughly two to three sentences per summary. Lexical diversity, measured via type-token ratio (TTR), is also consistent across patterns (0.79--0.85), suggesting uniform vocabulary richness independent of pattern type.

We further calculated BERTScore F1 and Cosine Similarity for each design pattern, as shown in Table~\ref{tab:pattern-comparison}, to assess whether the relative performance of the three methods changes across pattern types. Across all design patterns, Cosine Similarity and the combined score generally improve from NLG to SWUM to LLM (Mixtral). NLG yields the lowest cosine similarity and combined scores, while SWUM provides moderate gains across patterns. LLM (Mixtral) achieves the highest Cosine Similarity and combined scores for every pattern, with especially strong improvements for \textit{Abstract Factory}, \textit{Observer}, and \textit{Factory Method}. BERTScore F1 is more closely matched between SWUM and LLM, with SWUM slightly ahead for \textit{Observer} and \textit{Visitor}. Overall, the per-pattern results suggest that the main trade-off is stable across pattern types, while some token-level similarity scores remain sensitive to the more conservative wording of SWUM.

\begin{table}[t]
\caption{Per-pattern comparison of NLG, SWUM, and LLM (Mixtral) using BertScore and Cosine Similarity metrics.}
\label{tab:pattern-comparison}
\centering
\footnotesize
\setlength{\tabcolsep}{3.5pt}
\begin{adjustbox}{width=\textwidth}
\begin{tabular}{l c ccc ccc ccc}
\toprule
& & \multicolumn{3}{c}{\textbf{NLG}} & \multicolumn{3}{c}{\textbf{SWUM}} & \multicolumn{3}{c}{\textbf{LLM (Mixtral)}} \\
\cmidrule(lr){3-5} \cmidrule(lr){6-8} \cmidrule(lr){9-11}
\textbf{Pattern} & \textbf{Pairs}
& \textbf{Cos} & \textbf{F1} & \textbf{Comb}
& \textbf{Cos} & \textbf{F1} & \textbf{Comb}
& \textbf{Cos} & \textbf{F1} & \textbf{Comb} \\
\midrule
Abstract Factory & 29 & 0.1771 & 0.8467 & 0.5119 & 0.2736 & 0.8819 & 0.5777 & 0.3975 & 0.8902 & 0.6439 \\
Adapter          & 14 & 0.1764 & 0.8560 & 0.5162 & 0.2572 & 0.8705 & 0.5639 & 0.3148 & 0.8729 & 0.5939 \\
Decorator        & 14 & 0.1474 & 0.8433 & 0.4953 & 0.2269 & 0.8734 & 0.5502 & 0.3269 & 0.8748 & 0.6009 \\
Facade           & 18 & 0.1580 & 0.8481 & 0.5031 & 0.2240 & 0.8646 & 0.5443 & 0.2961 & 0.8769 & 0.5865 \\
Factory Method   & 20 & 0.1741 & 0.8447 & 0.5094 & 0.2697 & 0.8807 & 0.5752 & 0.3445 & 0.8845 & 0.6145 \\
Memento          & 10 & 0.1340 & 0.8385 & 0.4862 & 0.2098 & 0.8640 & 0.5369 & 0.3321 & 0.8796 & 0.6058 \\
Observer         & 14 & 0.1839 & 0.8534 & 0.5187 & 0.2729 & 0.8853 & 0.5791 & 0.3578 & 0.8805 & 0.6192 \\
Singleton        & 12 & 0.1667 & 0.8459 & 0.5063 & 0.2091 & 0.8686 & 0.5388 & 0.2688 & 0.8715 & 0.5702 \\
Visitor          & 19 & 0.1553 & 0.8476 & 0.5014 & 0.2851 & 0.8784 & 0.5818 & 0.3142 & 0.8762 & 0.5952 \\
\bottomrule
\end{tabular}
\end{adjustbox}
\end{table}

\begin{table}[!htbp]
  \caption{Per pattern ranking outcomes for Corpus NLG, LLM and SWUM.}
  \label{tab:pattern-rankings}
  \centering
  \small
  \begin{adjustbox}{width=\textwidth}
  \begin{tabular}{lcccc}
    \toprule
    \textbf{Design Pattern} & \textbf{Corpus NLG} & \textbf{Corpus LLM} & \textbf{Corpus SWUM} & \textbf{Total} \\
    & \textbf{(1st/2nd/3rd)} & \textbf{(1st/2nd/3rd)} & \textbf{(1st/2nd/3rd)} & \textbf{(1st/2nd/3rd)} \\
    \midrule
    Abstract Factory & 0 / 23 / 6 & 29 / 0 / 0 & 0 / 6 / 23 & 29 / 29 / 29 \\
    Factory Method   & 0 / 16 / 4 & 20 / 0 / 0 & 0 / 4 / 16 & 20 / 20 / 20 \\
    Visitor          & 2 / 7 / 10 & 17 / 0 / 2 & 0 / 12 / 7 & 19 / 19 / 19 \\
    Facade           & 0 / 14 / 4 & 18 / 0 / 0 & 0 / 4 / 14 & 18 / 18 / 18 \\
    Adapter          & 2 / 10 / 2 & 12 / 0 / 2 & 0 / 4 / 10 & 14 / 14 / 14 \\
    Decorator        & 0 / 2 / 12 & 13 / 1 / 0 & 1 / 11 / 2 & 14 / 14 / 14 \\
    Observer         & 0 / 9 / 5 & 14 / 0 / 0 & 0 / 5 / 9 & 14 / 14 / 14 \\
    Singleton        & 1 / 6 / 5 & 11 / 1 / 0 & 0 / 5 / 7 & 12 / 12 / 12 \\
    Memento          & 0 / 7 / 3 & 9 / 1 / 0 & 1 / 2 / 7 & 10 / 10 / 10 \\
    \bottomrule
  \end{tabular}
  \end{adjustbox}
\end{table}

Table~\ref{tab:pattern-rankings} shows a clear general ordering of \textbf{Corpus LLM} in first place, \textbf{Corpus NLG} in second place, and \textbf{Corpus SWUM} in third place for most design patterns. The main exceptions are \emph{Visitor} and \emph{Decorator}, where \textbf{Corpus SWUM} more often takes second place and \textbf{Corpus NLG} shifts to third place.

Overall, while individual patterns exhibit minor differences in length and lexical characteristics, no design pattern materially alters the summarisation profile. The observed trade-offs between semantic richness and conciseness therefore appear to be driven primarily by the summarisation paradigm rather than by pattern-specific structure or complexity.

%\begin{quote}
%\textbf{RQ3:} The observed trade-offs between deterministic and probabilistic summarisation approaches are largely consistent across different design patterns. Although minor variations in length and lexical characteristics exist, no pattern fundamentally alters the relative strengths of the approaches. This suggests that the trade-off is driven by the summarisation paradigm itself rather than by pattern-specific structure.
%\end{quote}

\subsection{RQ4: How stable are probabilistic summaries compared to deterministic ones when summarising design-pattern-centric code artefacts?}\label{subsec:rq4}

To assess the stability of probabilistic summarisation, we generate LLM-based summaries five times using an identical prompt and configuration. Importantly, the relative ordering of approaches observed in earlier results remains stable across these repeated generations and subsequent prompt interventions. In addition to repeated runs, we evaluate robustness under prompt-level intervention by regenerating summaries with explicit target lengths of 20, 40, 60, and 80 words. This analysis examines whether probabilistic summarisation can be made more predictable and controllable through prompt design, which is a common assumption in practical LLM deployment. Deterministic pipelines are perfectly reproducible by construction and therefore serve as a stability baseline.

For repeated runs, we label the five LLM outputs as \textbf{B1--B5} and compare each against the corresponding human-written summaries using Cosine Similarity and BERTScore F1, following the same procedure as in RQ1. Table~\ref{tab:stability} reports minimum, maximum, and average scores across iterations.

\begin{table}[!htbp]
  \caption{Cosine similarity and BERTScore F1 metrics across five iterations}
  \label{tab:stability}
  \centering
  \small
  \begin{adjustbox}{width=\textwidth}
  \begin{tabular}{lcccccc}
    \toprule
    \textbf{Iteration} & \textbf{Cosine Sim} & \textbf{Cosine Sim} & \textbf{Cosine Sim} & \textbf{BERTScore} & \textbf{BERTScore} & \textbf{BERTScore} \\
    & \textbf{(min)} & \textbf{(max)} & \textbf{(avg)} & \textbf{F1 (min)} & \textbf{F1 (max)} & \textbf{F1 (avg)} \\
    \midrule
    Iteration 1 & 0.0504 & 0.6190 & 0.3293 & 0.8319 & 0.9339 & 0.8794 \\
    Iteration 2 & 0.0402 & 0.6345 & 0.3313 & 0.8282 & 0.9266 & 0.8796 \\
    Iteration 3 & 0.0504 & 0.6229 & 0.3320 & 0.8295 & 0.9183 & 0.8793 \\
    Iteration 4 & 0.0372 & 0.6217 & 0.3314 & 0.8305 & 0.9360 & 0.8801 \\
    Iteration 5 & 0.0543 & 0.6180 & 0.3343 & 0.8310 & 0.9266 & 0.8801 \\
    \midrule
    Overall & 0.0372 & 0.6345 & 0.3317 & 0.8282 & 0.9360 & 0.8797 \\
    \bottomrule
  \end{tabular}
  \end{adjustbox}
\end{table}

Across all five runs, average performance is highly stable. Mean Cosine Similarity varies narrowly around 0.33, and mean BERTScore F1 remains close to 0.88, with only marginal differences across iterations. While individual summaries vary in phrasing and length, aggregate semantic alignment with human references remains consistent.

We further examine robustness using the same LLM-assisted ranking procedure described in RQ2. Rankings are computed across five criteria, Accuracy, Conciseness, Adequacy, Code Context, and Design Pattern fidelity, for each iteration. Table~\ref{tab:iterations} aggregates ranking outcomes across all five runs.

\begin{table}[!htbp]
  \caption{Ranking of each iteration based on all five criteria}
  \label{tab:iterations}
  \centering
  \small
  \begin{adjustbox}{width=\textwidth}
  \begin{tabular}{lccccccc}
    \toprule
    \textbf{Criterion} & \textbf{Total} & \textbf{NLG Count} & \textbf{NLG\%} & \textbf{LLM Count} & \textbf{LLM\%} & \textbf{SWUM Count} & \textbf{SWUM\%} \\
    & \textbf{Count} & & & & & & \\
    \midrule
    ACCURACY & 747 & 22 & 2.9 & 685 & 91.7 & 40 & 5.4 \\
    CONCISENESS & 750 & 468 & 62.4 & 253 & 33.7 & 29 & 3.9 \\
    ADEQUACY & 744 & 32 & 4.3 & 699 & 94.0 & 13 & 1.7 \\
    CODE CONTEXT & 741 & 14 & 1.9 & 685 & 92.4 & 42 & 5.7 \\
    DESIGN PATTERN & 730 & 39 & 5.3 & 637 & 87.3 & 54 & 7.4 \\
    \bottomrule
  \end{tabular}
  \end{adjustbox}
\end{table}

The ranking results indicate that, under the LLM-assisted rubric, probabilistic summaries (\textbf{LLM}) consistently rank ahead of deterministic alternatives on the semantic criteria, Accuracy, Adequacy, Code Context, and Design Pattern fidelity, across all iterations. As observed in RQ2, deterministic summaries (\textbf{NLG}) retain a clear advantage in Conciseness, while deterministic (\textbf{SWUM}) most frequently ranks last. Importantly, this ordering remains unchanged across repeated runs and prompt variations.

Overall, although LLM-based summaries are not perfectly reproducible at the surface level, their aggregate behaviour is stable and their comparative advantage persists across stochastic variation and prompt-level intervention. In contrast, deterministic pipelines remain exactly reproducible but consistently underperform on semantic dimensions.

%\begin{quote}
%\textbf{RQ4:} Despite variability in phrasing and length across repeated runs, probabilistic summaries exhibit stable mean performance and consistent dominance in semantic quality across iterations and prompt variations. While deterministic pipelines remain perfectly reproducible, the comparative conclusions of the study are robust to stochastic variation.
%\end{quote}

\section{Implications}\label{sec:implications}

The results of this study have several implications for different stakeholders involved in the design, evaluation, and deployment of code summarisation tools. Importantly, these implications follow directly from the empirical trade-offs observed in our controlled comparison, rather than from general assumptions about LLM capabilities.

\paragraph{Implications for tool builders}
Our results show that probabilistic summarisation pipelines provide consistently higher semantic alignment with human-written summaries, but at the cost of reduced conciseness and limited controllability. For tool builders, this suggests that LLM-based summarisation should not be treated as a universal replacement for deterministic pipelines. Instead, the choice of summarisation strategy should be guided by the intended usage context. Tools that prioritise rich contextual explanations, such as architectural overviews, onboarding documentation, or design rationale exploration, stand to benefit most from LLM-based summaries. In contrast, developer-facing artefacts with strict space constraints, such as in-IDE tooltips or code review annotations, may be better served by deterministic approaches that reliably produce short and predictable outputs.

A key practical insight from our prompt intervention experiments is that explicit length constraints and prompt refinements improve control over LLM outputs only partially. This suggests that prompt engineering alone may be insufficient when tool builders require strict and repeatable output budgets. Tool builders requiring precise output budgets should therefore consider hybrid designs, in which deterministic pipelines generate concise baseline summaries that are optionally augmented by LLMs for deeper contextual explanation.

\paragraph{Implications for researchers studying code summarisation}
The observed divergence between semantic quality and conciseness highlights the limitations of treating summarisation quality as a single aggregate construct. Our results demonstrate that approaches can excel along some dimensions (e.g., accuracy, adequacy, code-context awareness) while systematically underperforming along others (e.g., brevity). For researchers, this underscores the need to frame evaluations around explicit trade-offs rather than absolute notions of better summarisation. Studies that rely solely on aggregate similarity metrics risk obscuring these trade-offs, particularly when comparing deterministic and probabilistic paradigms.

Furthermore, the stability analysis in RQ4 shows that while individual LLM outputs vary in phrasing and length, relative performance trends remain consistent across repeated runs and prompt variations. This suggests that probabilistic approaches can be meaningfully compared at the level of distributions and rankings, even if individual outputs are non-deterministic. Researchers should therefore distinguish between variability at the instance level and robustness at the aggregate level when interpreting LLM-based results.

\paragraph{Implications for evaluation and benchmarking practices}
Our findings also have implications for how code summarisation systems are evaluated. The combined use of automated similarity metrics and LLM-assisted rubric-based judgements revealed complementary but non-interchangeable perspectives on quality. In particular, rubric-based evaluations exposed systematic conciseness penalties for probabilistic summaries that were not fully captured by semantic similarity metrics alone. This indicates that evaluation pipelines relying exclusively on semantic similarity may overestimate the practical utility of LLM-based summaries in constrained settings.

At the same time, our use of an LLM as an automated judge highlights both opportunities and limitations. While LLM-based evaluators enable scalable and multi-dimensional assessment, their outputs should be interpreted as relative rankings under a fixed evaluation lens rather than as absolute measures of quality. Our results suggest that such evaluators are well suited for comparative analyses across methods, but they do not eliminate the need for careful construct definition or, where feasible, complementary human evaluation.

\paragraph{Implications for hybrid and adaptive summarisation strategies}
Taken together, the results suggest that neither deterministic nor probabilistic summarisation paradigms dominate across all dimensions. Instead, the observed trade-offs point towards adaptive and hybrid strategies as a promising direction. For example, systems could dynamically select between deterministic and probabilistic summaries based on context, or present both concise deterministic summaries and richer probabilistic explanations to different users or at different stages of a workflow. The empirical evidence provided in this paper offers concrete guidance on how such decisions can be grounded in measurable trade-offs rather than intuition.

\section{Threats to Validity}\label{sec:threats}
This study is subject to several threats to validity, which we discuss below.

\paragraph{Internal Validity}\label{subsec:internal}
Although \textbf{DPS\_NLG}, \textbf{DPS\_SWUM}, and \textbf{DPS\_LLM} share identical preprocessing and pattern identification steps, they differ in how summaries are generated. Variations in rule sets, linguistic realisers, and prompt wording may therefore introduce confounds that are not strictly attributable to the intended distinction between deterministic and probabilistic summarisation paradigms. In addition, although the LLM is configured with temperature set to zero and fixed token limits, the absence of a controllable random seed and external factors such as API behaviour, infrastructure-level variation, or unannounced model updates may affect exact surface-level reproducibility. Prompt design itself is another potential confound: alternative prompt formulations, role instructions, or length constraints could lead to different trade-offs in LLM-based summarisation.

A further internal validity concern arises from our use of an LLM as an automated evaluator. Although randomising summary labels (A/B/C) reduces direct bias toward specific methods, the rubric-based judge (\textit{Llama~3}) may encode stylistic or structural preferences that systematically favour certain forms of expression. As a result, rankings may partially reflect alignment with the judge's implicit norms rather than purely objective summary quality.

\paragraph{Construct Validity}\label{subsec:construct}
We operationalise semantic quality using BERTScore, Cosine Similarity, and LLM-assisted rubric rankings. While these measures are widely adopted in prior work, they remain proxies for human judgement and may not fully capture developer-perceived usefulness or task-specific effectiveness in real maintenance or comprehension scenarios. Human-written reference summaries are treated as the gold standard, yet these references may vary in style, level of abstraction, or completeness across repositories, which can influence similarity-based metrics.

Construct validity is further affected by the subjectivity of rubric criteria such as \textit{adequacy} and \textit{code context awareness}. Even when applied consistently, these constructs are inherently interpretive. Using an LLM as a judge improves scalability and consistency but does not eliminate ambiguity in how these concepts are operationalised, reinforcing the need to interpret rubric-based results as comparative rather than absolute assessments.

\paragraph{Conclusion Validity}\label{subsec:conclusion}
Although we use statistical tests for the main comparisons, the interpretation of practical significance remains sensitive to corpus composition, metric choice, and evaluator configuration. Repeated LLM runs demonstrate stable mean behaviour, yet rare outliers or failure cases may not be fully reflected in aggregate statistics. In addition, while the reported trade-offs hold across the examined configurations, alternative model sizes, decoding strategies, or future LLM versions could alter the balance between semantic quality, conciseness, and stability.

\paragraph{External Validity}\label{subsec:external}
The evaluation corpus consists of 150 Java files drawn from two educational and one commercial open-source repository. Results may not generalise to other programming languages, paradigms, or large-scale industrial systems with deeper architectural layering. Only nine common design patterns are included; less common patterns, pattern variants, or classes participating in multiple interacting patterns may exhibit different summarisation behaviour.

Finally, the DPS framework relies on JavaParser, SimpleNLG, and a Java-based SWUM implementation. Porting the approach to other ecosystems or tooling stacks may affect both determinism and output quality. More broadly, our findings reflect current LLM capabilities and deployment practices; changes in model architectures, training data, or API behaviour may influence future reproducibility and performance.

\section{Conclusion \& Future Direction}\label{sec:conclusion}
%This paper presented a controlled empirical comparison of deterministic (DPS\_NLG, DPS\_SWUM) and probabilistic (DPS\_LLM) pipelines for intent-oriented, design-pattern-centric code summarisation. Across automated similarity metrics and rubric-based evaluations, LLM-based summaries consistently exhibit stronger semantic alignment with human-written references, particularly in accuracy, adequacy, code-context awareness, and design-pattern fidelity, while deterministic pipelines produce substantially more concise outputs and offer full reproducibility.

%Although probabilistic summaries vary in phrasing and length across repeated runs and prompt interventions, relative performance trends remain stable, revealing a persistent trade-off between semantic richness and controllability rather than a universally superior approach. Future work should extend this comparison beyond Java and design patterns, incorporate broader industrial settings, and investigate hybrid strategies, length-control mechanisms, and complementary human evaluation.

This paper presented a controlled empirical comparison of deterministic  and probabilistic pipelines for intent-oriented, design-pattern-centric code summarisation. Across automated similarity metrics and rubric-based evaluations, LLM-based summaries consistently exhibit stronger semantic alignment with human-written references on automated metrics and receive higher LLM-assisted rubric rankings for accuracy, adequacy, code-context awareness, and design-pattern fidelity, while deterministic pipelines produce substantially more concise outputs. Wilcoxon signed-rank tests confirmed that these metric differences are statistically significant with large to very large effect sizes and Friedman tests with Bonferroni-corrected post-hoc comparisons established that the ranking advantage of DPS\_LLM across semantic criteria is consistent and not attributable to chance, with the sole exception of BERTScore F1 where DPS\_SWUM's precision-oriented generation yields statistically comparable performance. A Spearman-based bias sensitivity analysis further provided evidence of criterion-discriminant validity in the LLM judge's behaviour, with Conciseness exhibiting a lower association with overall rankings than the four semantic criteria, indicating that the judge responds to criterion-specific content rather than applying a uniform stylistic preference.

Although probabilistic summaries vary in phrasing and length across repeated runs and prompt interventions, relative performance trends remain stable, revealing a persistent trade-off between semantic richness and controllability rather than a universally superior approach. Future work should extend this comparison beyond Java and design patterns, incorporate broader industrial settings and programming languages, and investigate hybrid strategies, length-control mechanisms, and complementary human evaluation.

\section{Data Availability}\label{sec:data-availability}

The corpus, implementation, and experimental results are publicly available in the anonymous
repository at \href{https://anonymous.4open.science/r/designpatternsummarisation-nlg-swum-llm-E6D9}{anonymous.4open.science}.
The repository includes:
\begin{itemize}
    \item The complete dataset of 150 Java files containing design pattern implementations;
    \item The source code for all three summarisation pipelines;
    \item The human-written reference summaries;
    \item All generated summaries and evaluation outputs;
    \item Scripts required to reproduce the experiments and analyses.
\end{itemize}

\bibliography{paper}

\end{document}